\documentclass{llncs}

\usepackage{bm}

\newcommand{\eqn}[2]
{
\begin{equation}#1
#2
\end{equation}
}

\newcommand{\eqna}[1]
{
\begin{eqnarray}
#1
\end{eqnarray}
}

\newcommand{\tab}[4]
{
\begin{table}[htbp]
  \centering
  \caption{#2}
  #1
  \begin{tabular}{#3}
  #4
  \end{tabular}
\end{table}
}

\newcommand{\sn}[1]
{
\section{#1}
}

\newcommand{\ssn}[1]
{
\subsection{#1}
}

\newcommand{\sssn}[1]
{
\subsubsection{#1}
}

\newcommand{\ra}
{
\rightarrow
}

\begin{document}

\title{Ranking Entity Based on Both of Word Frequency and Word Sematic Features}

\author{Xiao-Bo Jin\inst{1} \and Guang-Gang Geng\inst{2} \and Kaizhu Huang\inst{3} \and Zhi-Wei Yan\inst{4}}
\institute{Henan University of Technology  \email{xbjin9801@gmail.com} \and China Internet Network Information Center \and Xi¡¯an Jiaotong-Liverpool University \and National Engineering Laboratory for Naming and Addressing Technologies}

\maketitle

\abstract{Entity search is a new application meeting either precise or vague requirements from the search engines users. Baidu Cup 2016 Challenge just provided such a chance to tackle the problem of the entity search. We achieved the first place with the average MAP scores on 4 tasks including movie, tvShow, celebrity and restaurant. In this paper, we propose a series of similarity features based on both of the word frequency features and the word semantic features and describe our ranking architecture and experiment details.
}

\sn{Introduction}
The extraction of the feature vectors from the query-document is a critical step in learning to ranking. The main effort lies in mapping the query-document pair into a joint feature space which can precisely establish their relevance.

The common method is to extract the features for each of the input text pair and then rank them on the various similarity measures based on the lexical and semantic analysis. But the similarity defined in a different measure will lead to the different rank sequences. Surdeanu et al. \cite{surdeanu_learning_2011} explore a wide range of classes of the features such as coarse word sense disambiguation, name-entity identification, syntactic parsing, and sematic role labeling. Although there is a large mount of text resource in the Web, but the labeled semantic resource for the supervised learning such as Penn Treebank is rare especially for the minority language, e.g. Chinese or Korean.

Deep learning especially the Convolutional Neural Networks (CNN) has been recently shown that it can efficiently learn to embed the sentences into a low dimensional vector space but preserve their syntactic and semantic relations in many NLP tasks \cite{kim_convolutional_2014,zhang_character-level_2015}. Severyn and Moschitti \cite{severyn_learning_2015} build the CNN on the sentences pair in an end-to-end manner, where their model on the TREC question answering task outperforms the state-of-art systems without the manual feature engineering and the additional syntactic parsers.

In Baidu Cup 2016 Challenge, the competition invites the participants to tackle the problem of the Chinese entity search on four tasks including restaurants, movies, TV shows and celebrities. Given a query on the entity and a set of candidate entities, the ranking system should rank the entities with their relevance to the query. It is similar to the question answering, but with one characteristic: the answering only contains the entity name, which is too short to express its hiding meanings. For example, when querying "President of U.S.A.", can we predict the relevance of the celebrity 'Isaac Newton' to the question without domain knowledge? It is impossible to achieve a good performance without the domain knowledge if we only focus on the lexical, the statistical, and the semantical feature information of the answer.

In this paper, we describe a novel ranking learning architecture for the ranking of the entity object to help us to achieve the champion of Baidu Cup 2016 challenges. The distinctive properties of our architecture are: (1) we crawl the domain knowledge automatically to extend the answers according to the different tasks; (2) the statistical relevance features are extracted to build ranking models, but we only use the simple frequency features since the syntax or the semantical resource on Chinese text is difficult to obtain; (3) we also define the new semantic relevance features by means of word2vec to handle the large scale of corpus, in contrast to the CNN \cite{severyn_learning_2015} which is restricted to the short texts or sentences; (4) we adopt the simple point-wise method to take the relevance features as the input instead of directly computing their similarity.

We validate our architectures on the four tasks and analyze the effects of the components on the ranking performance, e.g. MAP or MRR. In the following, we give the formulation of the ranking problem and then describe the components of the ranking learning system and report our state-of-art experimental results. Finally, we conclude the paper and the outline the future work.

\sn{Backgrounds}
The section briefly describes entity search problem, then discusses learning to rank, which will train a rank model to predict the order of the candidate answer.

\ssn{Problem Formulation}
The query is a set of keywords or key-phrases used by the users to express their desire. An entity is a thing with distinct and independent existence such as celebrity, restaurant, movie and tvShow in our tasks.
The goal is to query the description targeting the entities, e.g. the review on some movie, the feeling in the restaurant environment.

Given an entity search query $q_i \in Q$ and a set of the candidate entities $E_i = {(d_1,r_1),(d_2,r_2),\cdots,(d_{i_k},r_{i_k}),(d_{i_n},r_{i_n})}$, where $d_{i_k}$ is the entity object and $r_{i_k}$ is the relevant label equal to 1 if relevant to the query and 0 otherwise. The objective is to retrieve entities that is relevant to the query $q_i$ from $E_i$ under the ranking function is:
\eqn{}{
h(\bm{w},\phi(q,D)) \ra R,
}
where $\phi(q,D)$ is a query-dependent features depending both on the entity and the query and $\bm{w}$ is the parameter of the ranking function.

\ssn{Learning to Rank}
The algorithms of learning to rank are traditionally classified as three categories. In the simplest point-wise approach, the instances are assigned a ranking score as the absolute quantity using
classical regression or classification techniques \cite{crammer_pranking_2001,li_mcrank:_2007}. McRank \cite{li_mcrank:_2007} casts the ranking problem as the classification or multiple ordinal regression classification, where Discount Cumulative Gain (DCG) errors are bounded by the classification errors. In the pairwise approach, the order of the instance pair is treated as a binary label and learned by a classification method. RankSVM \cite{joachims_optimizing_2002} formalizes this task as a problem of classifying  the instance pairs into two categories (correctly ranked and incorrectly ranked). RankBoost \cite{freund_efficient_2003} maintains $n$ weak ranking functions where each function can order the instances and then combines the ranking functions into a single ranking. LambdaRank optimizes the IR measure by defining a virtual gradient on the sorted document. Finally, the most complex list-wise approaches try to directly optimize a ranking-specific evaluation metric (e.g. NDCG). It often applies the continuous approximation or the bound of the evaluation measure since most evaluation measures are not a continuous function with respect to the parameters of the rank model
\cite{valizadegan_learning_2009,burges_learning_2006}. The report \cite{olivier_yahoo!_2011} on given by Yahoo Learning to Rank Challenge shows that nonlinear models such as the trees and the ensemble learning methods are powerful techniques and there is no significant difference among the above three kinds of methods on the large scale of the dataset by the extensive experiments. In our framework we adopt the simple point-wise method to predict the probability of the entity relevant to the query on the statistical and the semantic representation of the query-entity pairs.

\sn{Framework of Learning to Rank Entity}

\ssn{Extending Entity Extension by External Resources}

We have recently seen a rapid and successful growth of Baidu Baike \footnote{http://baike.baidu.com/}, which is a largest open Chinese encyclopedia on the Web. It has now more than 13,000,000 word-items edited by approximate 6,000,000 free volunteers or professional personnel. The Baike aims to be a Chinese encyclopedia and the articles on the Baike is refer to all aspects of the Chinese culture. We extract the knowledge for each entity object from celebrity, movie and tvShow. Baike will be much easier than from raw texts or from usual Web texts because of its structure. The objectiveness of the Baike also help us to rank the entity according to the query precisely. In fact, many natural language processing studies try to exploit Wikipedia as a knowledge source \cite{zesch_analyzing_2007,kazama_exploiting_2007} for the English language.

The restaurant tasks is an exception since there is no needs for each restaurant for the Baike which aims to provide the authoritative resources or knowledge. So we crawled the review pages from Dazhong review \footnote{https://www.dianping.com}, which is the largest city life website guiding the mass consumption in China.

Finally, we also collected the information and the user reviews for the movie and the tvShow tasks from Douban website \footnote{https://www.douban.com/}, which provides the information and reviews on the books, the movies (including tvShow) and the music generated by over 2 billions Chinese users.

\ssn{Feature for Learning}

In this section, we introduce the features extracted in our experiments which can be directly used by the learning algorithm. Each row of the matrix corresponding to the feature file represents a query-entity pair. The other files for each query-entity pair separately records the query id, the entity id, the answer id in the query and the label shows the entity is relevant to the query or not.

\tab{\label{tab:features}}{Statistical features on corpuses: the words streams come from the title, the body and the title + the body and it can be segmented phrase or 2-ngram words}{cl}{
\hline
No & Description \\
\hline
1 & Sum of TF of the query in the stream \\
2 & Sum of IDF of the query in the stream \\
3 & Sum of TFIDF of the query in the stream \\
4 & Sum of BM25 of the query in the stream \\
5 & Sum of LMIR.JM of the query in the stream\\
6 & Sum of LMIR.DIR of the query in the stream\\
7 & Sum of LMIR.ABS of the query in the stream \\
\hline
8 & Max of all SS distance in the stream \\
9 & Max of all SWS distance in the stream \\
10 & Max of all MS distance in the stream \\
11 & Max of all MWS distance in the stream \\
12 & Average of all SS distance in the stream \\
13 & Average of all SWS distance in the stream \\
14 & Average of all MS distance in the stream \\
15 & Average of all MWS distance in the stream \\
\hline
}

\sssn{Word Frequency Features}
In the following, We give some details of these features, the fore part of which is referenced to the dataset LETOR 3.0 \cite{qin_letor:_2010}.

In the corpus, we considered three types of streams: title, body and title + body. For the entity body, we cut the whole part into the sentences by the end mark of the Chinese language and the English language for convenience. We also removed all Chinese and English punctuation character.

The term frequency of each query term is the count that it appeared in the stream. We separately computed the IDF (inverse document frequency) for every stream (e.g. title, body) by the following formula
\eqn{}{
idf(q_i) = \log\frac{N - n(q_i) + 0.5}{n(q_i) + 0.5}
}
where $N$ is the total count of the entities in the corpus and $n(q_i)$ is the frequency of the query $q_i$ in the stream. Finally, we sum the TF value, the IDF value and the TF value weighted by IDF of each query word (See 1-3 in Table \ref{tab:features}).

Okapi BM25 is a ranking function used to rank the documents according to their relevance to a given search query. For the query, its BM25 score (See 4 in Table \ref{tab:features}) is computed by
\eqn{}{
BM25(q,d) = \sum_{q_i:f(q_i,d) > 0} idf(q_i)\cdot \frac{f(q_i,d) \cdot (k_1 + 1)}{f(q_i,d) + k_1\cdot(1 - b + b \frac{|d|}{avg(d)})} \cdot \frac{(k_3 + 1)f(q_i,q)}{k_3 + f(q_i,q)}
}
where $f(q_i,d)$ is the times of $q_i$ occurring in the document $d$, $f(q_i,q)$ is the times of $q_i$ occurring in the query $q$, $|d|$ is the count of the words (or the length of the document) in the document $d$ and $avg(d)$ is the average document length in the entire corpus. According to the paper \cite{qin_letor:_2010}, we empirically set $k_1 = 2.0$, $k_3 = 0$ and $b = 0.75$.

LMIR.JM, LMIR.ABS and LMIR.DIR in Table \ref{tab:features} come from the language model related features, we follow along the line of the paper \cite{zhai_study_2001}. With a uniform prior, the language model reduces to the calculation of $p(q|d)$
\eqn{}{
\log p(q|d) = \prod_{i} p(q_i|d),
}
where $p(q_i|d)$ can be estimated by any language model mentioned by the above. The Jelinek-Mercer (LMIR.JM) implements a linear interpolation of the maximum likelihood model with the corpus model
\eqn{}{
p(q_i|d) = (1 - \lambda) p_{ml}(q_i|d) + \lambda p(q_i|C)
}
where $\lambda$ is set to $0.1$ to control the influence of the model, $p_{ml}(q_i|d)$ and $p(q_i|C)$ is the document probability and the corpus probability of $q_i$ estimated by the term frequency. Bayesian smoothing using Dirichlet priors (LMIR.DIR) is a multinomial distribution with the Dirichlet prior
\eqn{}{
p(q_i|d) = \frac{f(q_i,d) + \mu p(q_i|C)}{\sum_{q_i \in q} f(q_i,d) + \mu}
}
where $\mu$ is set to $2000$. Absolute discounting discounts the seen word probability by subtracting as constant instead of multiplying it by $(1 - \lambda)$ like LMIR.JM
\eqn{}{
p(q_i|d) = \frac{\max(f(q_i,d) - \delta,0) + \delta |d|_{\mu} p(w|C)}{\sum_{q_i \in q}f(q_i,d)},
}
where $|d|_{\mu}$ is the number of unique terms in the document $d$ and $\delta$ is set to $0.7$.

\sssn{Word Semantic Features}
Word2vec \cite{mikolov_efficient_2013} is a series of models used to produce the word embeddings, where the models are a two-layer neural networks that take as the input a large corpus of text and produce a corresponding vector for each unique word in the high-dimensional space. Although word2vec plays a part just for computing the similarity between the words, we have no knowledge about the computation of the sentence similarity. In the following, we first give some heuristic approaches to compute the similarity between any sentence based on the public available word2vec \footnote{https://code.google.com/archive/p/word2vec/}.

The similarity between the query word $q_i$ and the sentence $s$ is defined as the max value among the similarity between $q_i$ and the word $s_i$ in the sentence $s$
\eqn{}{
sim(q_i,s) = \max_{s_j \in s} q_i^T s_j,
}
where $q_i$ and $s_i$ is a normalized vector with the unit length and both of them are extracted on the trained model from all types of entity corpus by word2vec.

We arrange all $q_i \in q(i = 1,2,\cdots,m)$ into the matrix $Q = [q_1,q_2,\cdots,q_m]^T$ and $s_j \in s(j = 1,2,\cdots,n)$ into the matrix $S = [s_1,s_2,\cdots,s_n]^T$, then
\eqn{}{
R = QS^T
}
where $R = [r_1,r_2,\cdots,r_m]^T$ and $r_i = q_i^T s$. It is clear that
\eqn{}{
sim(q_i,s) = \|r_i\|_{\infty}
}

The similarity computation of the query $q$ and the sentence $s$ is related to $sim(q_i,s)$ for all $q_i \in q$. With the sum and max operation, we can define the following four features including Sum of Similarity (SS), Sum of Weighted Similarity (SWS), Max of Similarity (MS) and Max of Weighted Similarity (MWS)
\eqna{
SS(q,s) & = & \sum_{q_i \in q} sim(q_i,s) \\
SWS(q,s) & = & \sum_{q_i \in q} sim(q_i,s)*idf(q_i) \\
MS(q,s) &  = & \max_{q_i \in q} sim(q_i,s) \\
MWS(q,s) & = & \max_{q_i \in q} sim(q_i,s)*idf(q_i)
}

\ssn{Word Segmentation and 2-Gram Words}
Chinese Word segmentation is the problem of dividing a string into its component words. It is a critical step for Chinese language processing. But the performance of the algorithm depends the domain specific dict and the used corpus. The most word segmentation algorithm does not handle the ambiguous words and unregistered ones. In our work, we adopt the simple 2-gram representation to complement the deficiencies of the word segmentation considering most of Chinese phrases consist of two words.

\ssn{Classifier Design}
In our work, we chose the ensemble approaches to predict the similarity probability for each query-entity pair. Our experiment took three candidates including AdaBoost, Random Forest and ExtraTree Classifier \cite{pedregosa_scikit-learn:_2011}. Further, we also tried to fuse the posterior probabilities and the rankings from multiple classifiers although the improvement is subtle on the celebrity and the restaurant datasets.

\sn{Experiments}

\ssn{DataSet Description}
Baidu Challenge 2016 includes four datasets including movies, tvShows, restaurants and celebrities. For each type, there are 100 entity queries for the training and 1,000 ones for the testing. Before the competition ends, 40\% of the test queries were used as the development set for all participants. In our experiments, we do not consider the results on the development set. We extracted the feature vectors from all query-entity pairs, where Tab. gives the detailed information.

\tab{\label{tab:dataset}}{DataSet Information}{lrrrr}{
\hline
Name & \#Training & \#Training & \#Test & \#Test \\
     &  Queries   &   Examples  &  Queries  & Examples \\
\hline
movie &  100 & 9,596 &  1,000 & 98,309 \\
tvShow & 100 & 10,264 & 1,000 & 103,409 \\
celebrity & 100 & 9,939 & 1,000 & 99,785\\
restaurant & 100 & 9,983 & 1,000& 99,796 \\
\hline
}

\ssn{Experiments Design and Results}

\sssn{Data Retrieval}
We crawled Baike, Douban, Dazhong web sites by the Baidu crawler and Yahoo crawler. It is important to validate the correctness of the crawled pages. We saved the meta information in the front of the texts to check whether it is consistent with the corresponding entity or not.

\sssn{Preprocessing}

We preprocessed the Chinese texts by removing all Chinese punctuations after splitting the total texts into the sentences with the Chinese punctuations as the end mark. Further, the sentences were split into the single Chinese words by the Jieba open source \footnote{https://github.com/fxsjy/jieba} and  by sliding on the text with the two-width window (2-gram), separately.

\sssn{Word Embeddings} We initialized the word embeddings by running word2vec tool \cite{mikolov_distributed_2013} on the Chinese corpus. The tvshow and movie tasks used the corpus contain roughly 2 million vocabularies, 1 million ones for the celebrity tasks and 0.8 million ones. To train the embeddings we used the continuous of bag words model with the window size 5 to generate a 50-dimensional vectors for each word. The embeddings vector not present in the word2vec model were randomly initialized with the equal length vector with each component taken from the uniform distribution $U[-0.25,0.25]$. On both of word segmentation form and 2-ngram form, we extracted 33 features from each corpus and merged them into 66-dimensional features.

\sssn{Experiment Setup}
We evaluated the performance on the training dataset by 10-fold cross validation (cv-10). In particular, the query-entity pair from the same query would be put into the same fold for keeping the completeness of each query. The cross folds were kept invariant for all parameters settings.

In implementing the extra tree classifier, the parameter $n\_estimator$ were randomly drawn from the integer range $[100,500]$ and another parameter $max\_depth$ was randomly from the enumeration range $\{4,6,8,10,12\}$. The models parameters were optimized in the space of the grid with the parameter $n\_estimator$ and $max\_depth$ by the cross-validation on the training data.

\sssn{Evaluation Measures}
The competition uses the Mean Average Precision (MAP) to evaluate the quality of the submission file, which is common in the information retrieval. MAP examines the ranks of all the related entities and computes the mean over the average precision scores for each query
\eqn{}{
MAP(q) = \frac{1}{|q|} \sum_{q_i \in q} avgprec(q_i).
}
Meanwhile, we computes $avgprec(q_i)$ as follows
\eqn{}{
avgprec(q_i) = \sum_{e \in q_i}\frac{rel(e,q_i)}{pos(e)}
}
where $rel(e,q_i)$ (1 or 0) shows whether the entity $e$ is correlated with the query $q$ and $pos(e)$ is the position of the entity $e$ in the ranking sequence of the query $q$.

Finally, we achieve the first place of the Baidu Challenge 2016 competition as shown in the Tab. \ref{tab:results}. Limited to the short readiness time, more experiments and analysis are ongoing in order to keep the integrity of the entire paper. In the further work, we will promptly give more detailed comparisons and comprehensive experimental analysis.

\tab{\label{tab:results}}{Competition Results on Four Tasks}{lccccc}{
\hline
Tasks & Celebrity & Movie & Restaurant & TvShow & Total \\
\hline
Results & 0.8818& 0.7759& 0.5939&   0.5978&  0.7124 \\
\hline
}

\sn{Conclusion}
In this paper, we explore the merging of the simple word frequency features and the word2vec-based sematic features to solve the entity search problem. The effectiveness of the features is shown on the Baidu Challenge 2016 competitions datasets. We explain the entire processing of the experiments.  We achieved the best performance with the merging of the features and the extra tree ranker (point-wise ranker). In future work, we will improve and finish the experiment comparisons, furthermore, we will take these features as the preprocessing step of the deep learning and apply the CNN to learn the relation between the query and the entity.


\section*{Acknowledgment}
This work was partially supported by the Fundamental Research Funds for the Henan Provincial Colleges and Universities in Henan University of Technology, the National Basic Research Program of China (2012CB316301), the National Natural Science Foundation of China (61103138, 61005029,61375039 and 61473236).

\bibliographystyle{plain}

\end{document}